\documentclass[aps,prl,twocolumn,amsmath,amssymb,floatfix,superscriptaddress]{revtex4}
\usepackage{amsfonts}
\usepackage{amssymb}
\usepackage{graphicx}
\usepackage{bm}

\def\avg#1{\left\langle#1\right\rangle}

\def\be{\begin{equation}}       \def\ee{\end{equation}}
\def\bea{\begin{eqnarray}}      \def\eea{\end{eqnarray}}
\def\ba{\begin{array} }
\def\ea{\end{array} }
\def\bnum{\begin{enumerate} }
\def\enum{\end{enumerate}}

\def\=>{\Rightarrow}
\def\>{\rightarrow}

\def\PRB{Phys. Rev. B}

\def\eye2{Fathbb{I}}

\begin{document}
\title{Phases of the infinite U Hubbard model}
\author{Li Liu}
\affiliation{Department of Physics, Stanford University, Stanford, CA 94305}
\author{Hong Yao}
\affiliation{Department of Physics, University of California, Berkeley, CA 94720}
\affiliation{Materials Sciences Division,
Lawrence Berkeley National Laboratory, Berkeley, CA 94720}
\author{Erez Berg}
\affiliation{Department of Physics, Harvard University, Cambridge, MA 02138}
\author{Steven A. Kivelson}
\affiliation{Department of Physics, Stanford University, Stanford, CA 94305}
\date{\today}
\begin{abstract}
 We apply the density matrix renormalization group (DMRG) to study the phase diagram of the infinite $U$ Hubbard model on 2-, 4-, and 6-leg ladders.  Where the results are largely insensitive to the ladder width, we consider the results representative of the 2D square lattice model.  We find a fully polarized ferromagnetic Fermi liquid phase when $n$, the density of electrons per site, is in the range $1>n>n_F \approx 4/5$.  For $n=3/4$ we find an unexpected commensurate insulating ``checkerboard'' phase with coexisting bond density order with 4 sites per unit cell and block spin antiferromagnetic order with 8 sites per unit cell.
 For $3/4 > n$, the wider ladders have unpolarized groundstates, which is suggestive that the same is true in 2D.
\end{abstract}
\maketitle

The Hubbard model is the paradigmatic representation of strongly correlated electron systems in condensed matter physics. It is widely invoked in studies of metallic ferromagnetism, unconventional superconductivity, various forms of charge and spin density wave formation, and even in theoretical explorations of spin-liquid and non-Fermi liquid states. These studies often involve conflicting claims concerning the phase diagram of this model in more than one dimension, especially when the interaction strength $U$ is comparable to or larger than the bandwidth.

\begin{figure}[b]
\includegraphics[width=3. in]{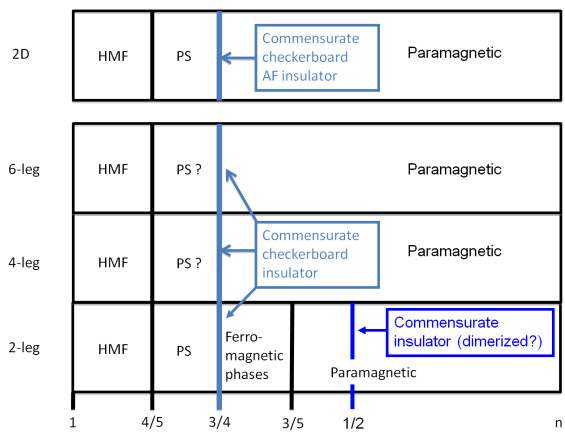}
\caption{The phase diagrams of the infinite $U$ Hubbard model on the
2-, 4-, and 6-leg ladders, and the inferred phase diagram in 2D.}
\label{fig:infer}
\end{figure}

In this paper we report the results of an extensive DMRG study of the zero temperature ($T=0$) phase diagram of the Hubbard model in the $U\to\infty$ limit, as a function of $n$,  the density of electrons per site.  To begin with, we study the two leg ladder on systems of size up to 2 $\times$ 50, large enough that finite size scaling can be used to obtain clear convergence to the thermodynamic limit.  The resulting phase diagram  is shown in the lower panel of Fig. \ref{fig:infer}.  It is frequently found that important features of two dimensional (2D) quantum models are already apparent in the solution of the 2-leg ladder.  To get a feeling for which features of the 2-leg Hubbard ladder extrapolate smoothly to 2D, we compute the properties of 4-leg and 6-leg ladders (with sizes up to 4 $\times$ 20 and 6 $\times$ 8, respectively).  The inferred partial phase diagrams of these wider ladders are shown in the two middle panels of Fig. \ref{fig:infer}.  (We have also carried out limited additional studies of 3 and 5 leg ladders.)  While there may be subtle correlations characteristic of the 2D model that would only be manifest were we able to study wider or longer ladders, many features of the phase diagram are already remarkably insensitive to ladder width and length for the studied system sizes.  We therefore speculate that these features survive as groundstate phases of the fully 2D model, as shown in the upper panel of Fig. \ref{fig:infer}.

To summarize our findings:

1) \underline {For $1 > n > n_F$}, we find
 a half-metallic ferromagnetic (HMF) phase, {\it i.e.} a fully spin polarized Fermi liquid.  For all even leg ladders we have studied $n_F =4/5$, so we expect that in 2D, $n_F \approx 4/5$ as well.  As is well known, the existence of a HMF phase for $n$ close enough to 1 is suggested by Nagaoka's theorem,\cite{nagaoka} and by various previous exact diagonalization\cite{young} and variational\cite{variational,recent} studies, the most recent of which\cite{recent} yields estimates of $n_F$ which are more or less consistent with the present results.  On the other hand, other lines of analysis\cite{tian} are suggestive that the HMF is stable only below a critical ``doped hole density,'' $
 (1-n)$, which vanishes in the thermodynamic limit ($n_F \to 1^-$); the present results falsify this.

2)  \underline {For $n_F > n > 3/4$}, the 2-leg ladder appears to phase separate (PS), with the two coexisting phases having densities $n=n_F$ and $n=3/4$.   Our more limited results on broader ladders suggest that the same might
hold true for 4 leg
ladders
and, by extension, in 2D as well.

3)  \underline {For $n=3/4$}, the 2-leg ladder forms an insulating commensurate plaquette density wave state, as shown 
 in Fig. \ref{qt} (a).  This pattern of symmetry breaking suggests that the spin-degrees of freedom correspond to an effective spin 3/2 Heisenberg antiferromagnetic chain on the ``checkerbord'' lattice defined by the ``strong'' plaquettes; indeed there is no detectable spin gap and we find clear signatures of quasi-long-range antiferromagnetic order with twice the period of the
plaquette order, as also illustrated in Fig. \ref{qt}(a). The 4-leg ladder exhibits a
similar (and slighly weaker) ordering tendency, forming the checkerboard plaquette order of the sort shown in Fig. \ref{qt}(b);  the spin correlations are consistent with those of an effective 2-leg spin 3/2 antiferromagnet on the checkerboard lattice in that there appears to be a small spin-gap, with the implication that, in this case, the corresponding antiferromagnetic order (also shown in the figure) decays exponentially at long distances.  Given that the 6-leg ladder (Fig. \ref{qt}(c)) also shows a similar ordering pattern as the 4 and 2-leg ladders, we are confident that the corresponding phase persists in the 2D limit.
The 2D ``checkerboard'' phase has coexisting bond-density wave and block-spin antiferromagnetic order\cite{yao}.  However, the site-charge density is uniform.  The existence of this phase, and its apparent robustness, was unanticipated in previous studies, as far as we know.

4)  \underline {For $3/4 > n > n_p$ with $n_p\approx 3/5$}, the 2-leg ladder has a partially polarized ferromagnetic ground-state, with maximal spin-polarization for $n\approx 2/3.$  We have not yet determined where, in this interval, the system forms a phase separated two-phase mixture and where it is in a pure phase. In any case, whatever the phase diagram of the 2-leg ladder, it is unlikely to be representative of the 2D system in this range of $n$, given that the groundstate is
paramagnetic for the 4-leg ladder.

5)  \underline {For $n_p > n$} (but with $n\neq 1/2$ and $n$ not too near 0), the groundstate of the 2-leg ladder is a paramagnetic
Luttinger liquid (LL) with   charge and spin gaps
that vanish in the thermodynamic limit to within
our numerical uncertainty.
The ground-states of the 3-leg and 4-leg ladders of all lengths we have studied are also spin-unpolarized in this range of $n$, from which we infer that the same is true of the 2D system.

6) \underline{For $n=1/2$} the 2-leg ladder exhibits a weakly dimerized insulating phase with a very small spin-gap.  However, we have no evidence that this insulating phase persists for wider ladders.

\noindent{\bf DMRG applied to Hubbard ladders:}
The Hubbard model is defined, as usual, by
\bea
H=-t\sum_{\avg{ij},\sigma=\uparrow,\downarrow} \left[c^\dag_{i\sigma}c_{j\sigma}+H.c.\right] +U\sum_ic^\dag_{i\uparrow}c_{i\uparrow} c^\dag_{i\downarrow}c_{i\downarrow},~~
\label{eq:a}
\eea
where $c_{j\sigma}^\dagger$ creates an electron with spin polarization $\sigma$ on site $j$ and $\avg{ij}$ signifies pairs of nearest-neighbor sites.  In the limit $U\to\infty$, the Hamiltonian is parameter-free in the sense that the second term in $H$ is replaced by the non-holonomic constraint of no double-occupancy:  $\sum_\sigma c_{j\sigma}^\dagger c_{j\sigma}= 0$, or 1, and one can chose units of energy such that $t=1$.

The DMRG calculations were carried out keeping up to 4000$\sim$18000 states. All ladders are taken to have open boundary conditions in both directions.  When we compute the expectation value of various densities, it is sometimes useful to break spin rotational symmetry by applying a Zeeman field of magnitude $h=1$ in the $z$ direction on a single boundary site, taken to be the site at the
lower left-hand end of the ladder.

To characterize the excitation spectrum of the system, we define the charge, spin, and single-particle gaps, $\Delta_c$, $\Delta_s$, and $\Delta_{1p}$, as follows:
\bea
\Delta_c\equiv &&[E(N_{el}+2) + E(N_{el}-2) - 2E(N_{el})]/2 \\
\label{gaps}
\Delta_s\equiv && [E(S=1;N_{el}) - E(S=0;N_{el})] \nonumber \\
\Delta_{1p}\equiv &&[E(N_{el}+1) + E(N_{el}-1) - 2E(N_{el})] \nonumber
\eea
where $N_{el}$ is the total number of electrons (which is always taken, for present purposes, to be even), and $E(N_{el})$ and $E(S,N_{el})$ are, respectively, the ground-state energy and the ground-state energy in a given spin-sector.  This definition of the spin-gap is only useful under circumstances in which the groundstate has $S=0$.  Where possible, we have extrapolated values of the gaps to the thermodynamic limit by fitting the data from finite length ladders to a quadratic form, $\Delta(N) = \Delta + A N^{-1} + B N^{-2}$, where $N$ is the length of the ladders;  unless otherwise stated, all gaps listed below have been extrapolated to the thermodynamic limit in this way.  Details of the various extrapolations will be presented at a future date.

Clearly, in a Fermi liquid (FL), all three gaps vanish in the thermodynamic limit.  In 1D, the FL is unstable in the presence of any interactions, but there exists a gapless LL which is a direct descendant of the FL.

\begin{figure}[b]
\includegraphics[width=3.5 in]{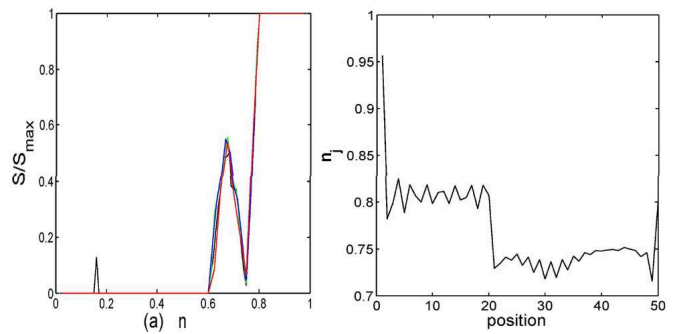}
\caption{(a) Magnetization, normalized to its
maximum possible value, of the 2-leg ladder as a function of $n$.The black, green, blue, and red curves correspond to N=50, 40, 30, and 20. (b) Expectation value of the site density, $n_j$, in a 2 $\times$ 50 ladder with average density $n=0.77$ and a chemical potential of magnitude $\mu=0.3t$ applied to the left-most 2 $\times$ 20 sites. 
}
\label{fig:pd}
\end{figure}

{\bf Results for the 2-leg ladder:}
We have computed the groundstate properties of the 2-leg ladder as a function of $n$ for system sizes 2 $\times\ N$ with $N=20$, 30, 40, and 50.
 To identify the ferromagnetic portions of the phase diagram, we have computed the ground-state magnetization density $ M = S/S_{max}$, where $S$ is the total spin of the groundstate, and $S_{max}=N n$ is the maximum possible value of $S$ in a fully spin-polarized state.  The results are shown in Fig. \ref{fig:pd} (a), where different colors denote
the different system sizes. Since
the four curves are nearly identical, the extrapolation to the thermodynamic limit is trivial.  Specifically:

\begin{figure}[b]
\includegraphics[width=3.5 in]{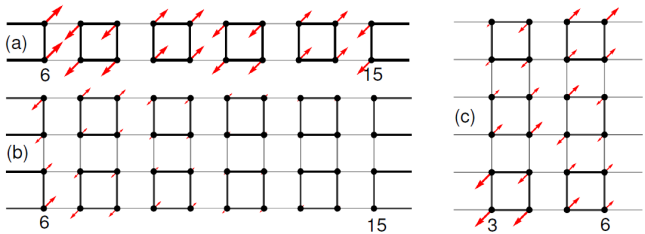}
\caption{Ground-state correlations at $n=3/4$ for  a central portion of (a) the  2 $\times$ 20 ladder, (b) the 4 $\times$ 20 ladder, (c) the 6 $\times$ 8 ladder.  The thickness of the lines is 
proportional to the third power of the magnitude of $B_{ij}$ 
and the length of the 
arrows to the magnitude 
of $S_{j}$, where the axis of spin-quantization is set by a Zeeman field of strength $h=1$ applied to the
lower left-hand site of each ladder. The numbers index the position along the ladder. For the 2 $\times$ 20 ladder, the values of $B_{ij}$ in the figure range from $B_{ij}= 0.30$ (lightest line) to $B_{ij} = 0.45$  (darkest line), while the magnitude of  $S_j$ ranges from -0.19 to 0.21. For the 4 $\times$ 20 ladder 
$B_{ij}$ ranges from $B_{ij}= 0.30$ to 0.41, while 
$S_j$ ranges from -0.12 to 0.12.  For the 6 $\times$ 8 ladder 
 $B_{ij}$ 
 ranges from $B_{ij}= 0.31$ to 0.41, while 
 $S_j$ ranges from -0.18 to 0.16. }.
\label{qt}
\end{figure}

1)The fully polarized ground state terminates at $n=n_F= 4/5$, independent of
$N$ \cite{liang}.  (See Table \ref{fm}.) This value of $n_F$ is not locked by any obvious commensurability effect that we have detected \cite{kohno}.  For instance, if we modify the Hamiltonian  by making the hopping matrix elements on the rungs, $t' = 0.5 t$, where $t$ is the hopping matrix element on the legs of the ladder, we find that $n_F = 0.85$. 

2) The ground state at $n=3/4$ is an insulating paramagnetic state with a charge gap, $\Delta_c=0.24 \pm 0.02 t$, a single-particle gap, $\Delta_{1p}= 0.245 \pm 0.02 t$, but a vanishing spin-gap
($\Delta_s <
3\times 10^{-4} t$ which is zero within our uncertainty). 
 The character of this state on a central segment of the ladder is shown in
Fig. \ref{qt}.  The thickness of the lines on the bonds between sites represents the 
magnitude of the expectation value of the bond-density,
\be
B_{ij}\equiv  \sum_\sigma\avg{\left[ c^\dagger_{i\sigma}c_{j\sigma} + {\rm h.c.}\right]}
\ee
and the length of the arrow on the site represents the expectation value of the spin on that site,
\be
S_j\equiv \frac12\sum_{\sigma}\sigma\avg{c_{j\sigma}^\dagger c_{j\sigma}}.
\ee
Spin-rotation invariance has been explicitly broken in this calculation by the application of a Zeeman field on the first site on the
lower leg of the ladder; translation symmetry is broken by the ends of the ladder.
There is a period-2 bond-density wave, with a magnitude that does not decrease with distance from the end of the ladder, nor does it depend significantly on the length of the ladder;  this signifies a discrete, broken translation symmetry.  There is also a period-4 ordering tendency of the spin density, but with an intensity which decays slowly with distance from the end at which the Zeeman field is applied.  This, and the absence of a spin-gap, signifies that there is quasi-long range antiferromagnetic order.
What is not shown in the figure is that the expectation value of the electron density, $n_j\equiv \sum_{\sigma}\avg{c_{j\sigma}^\dagger c_{j,\sigma}}$, 
is nearly uniform, $n_j\approx n$ for all $j$ except in the immediate vicinity of the ends of the ladder.

A way to visualize this state is that the ladder spontaneously forms a plaquette state consisting of a checkerboard array of weakly coupled plaquettes, each with three electrons in a state of total spin 3/2, and a weak, antiferromangetic exchange coupling between plaquettes.  As one would expect, the character of this state is not sensitive to small changes in the Hamiltonian -- for instance, if we set the rung hopping $t'=0.5t$ or $2t$, we find no qualitative change in the insulating phase at $n=3/4$.

3)  When $n$ is in the range $4/5 > n> 3/4$  the ground state appears to be a 2-phase mixture of the fully polarized state with $n=4/5$ and the checkerboard state with $n=3/4$. 
The evidence for this is as follows:  a) As seen in Fig. \ref{fig:pd} (a), the ground state magnetism
is (within expected finite size corrections) a linear function of $n$ in this range.
b) The groundstate energy (not shown) is, to similar accuracy, a linear function of $n$ in this range, with a continuous first derivative at $n=n_F$;  this is precisely the behavior expected from a Maxwell construction for a two-phase region.
 c)  As a final test,
 we have applied an on-site 
potential of magnitude $\mu=0.3t$ to the left-most
 2$\times$20 lattice sites of a 2$\times$50 ladder with mean value of $n=0.77$;
as can be seen in Fig. \ref{fig:pd} (b), the resulting density profile 
consists of a region with density $n_j\approx 4/5$ on the
left portion, and $n_j\approx 3/4$ on the right portion of the ladder, with a sharp domain wall separating them.

4)Again from Fig. \ref{fig:pd} (a), it is apparent that for $3/4 > n > 3/5$, there is a regime in which the ground-state is partially spin-polarized, with maximal spin-polarization being attained at $n=2/3$ where $M \approx 0.5$.  The behavior of the 2-leg ladder in this regime is interesting in its own right, but, in contrast to the situation in other ranges of $n$, similar behavior is not seen in wider ladders, as we shall see below.  We have thus not yet explored the nature of the phases that occur in this range of $n$ in any detail.

5)  For $3/5 > n$, the groundstate has $M=0$.  Finite size scaling leads to the
speculation that, for $n\neq 1/2$, this is a Luttinger liquid phase with  $\Delta_c$, $\Delta_s$ and $\Delta_{1p}$ all tending to 0 in the thermodynamic limit to within our numerical uncertainty of $\sim 0.02t$. In the near future we intend to perform more extensive studies of this regime to more fully explore the nature of the long-range correlations.

6) For $n=1/2$ there is a clearly identifiable charge gap,
 $\Delta_c \sim 0.1t$.  One can think of this as arising from a state in which there is a single electron localized on each rung of the ladder, so that the spin-degrees of freedom form an effective spin-1/2 chain, and thus can be expected to exhibit one of two possible phases -- a gapless phase with power law 
antiferromagnetic correlations 
or a dimerized phase with a spin-gap.  As we shall show in a forthcoming study, the 2-leg ladder with $n=1/2$ exhibits weak dimerization and fairly long-range antiferromagnetic correlations, implying that it is in the dimerized phase close to the quantum critical point.  Further evidence of this is provided by studying the changes produced by changing the ratio of the hopping matrix elements along the rungs and legs of the ladder, $t'/t$.  We find that the  dimerization is enhanced and the spin-correlations rendered short-range for $t'/t=0.5$, while the spin-correlations strengthened for $t'/t=2$.

\begin{table}
\begin{tabular}{|c|c|c|c|c|c|c|}
\hline
lattice size  & $n_F$& $ \epsilon_0$ & & lattice size  & $n_F$& $ \epsilon_0$\\
\hline
2$\times$20 &  0.800 & -0.6183& & 3$\times$20 &  0.867 &-0.4442\\
\hline
2$\times$30  & 0.800 &-0.6216& & 3$\times$25 &  0.867 & -0.4455\\
\hline
2$\times$40  & 0.800 & -0.6233& & 3$\times$30 &  0.867 & -0.4463\\
\hline
2$\times$50  & 0.800 & -0.6243& & 3$\times$35 &  0.867 & -0.4469\\
\hline
\hline
4$\times$10 &  0.800 & -0.6526& & 5$\times$25 & $0.827-0.833$ &  \\
\hline
4$\times$15  & 0.817 &-0.6089& & 5$\times$30 & 0.820 & -0.6558\\
\hline
4$\times$20  & 0.800 & -0.6613& &  & & \\
\hline
\hline
6$\times$10 &  0.817 & -0.6175& & 6$\times$20 &  0.808 & -0.6515\\
\hline
6$\times$15  & 0.800 & -0.6733& &  &   & \\
\hline
\end{tabular}
\caption{The
minimum electron density, $n_F$, for which the groundstate is fully polarized, $M=1$, for various size
ladders.  Here $\epsilon_0$ denotes the groundstate energy per site when $n=n_F$.}
\label{fm}
\end{table}

{\bf Results for wider ladders:}    Wider ladders provide necessary clues concerning the evolution of the phase diagram as the 2D limit is approached.  However, with increasing width, it becomes more difficult to obtain fully converged results from DMRG;  we are thus restricted to increasingly shorter ladders as the width increases.

    To study the evolution of the HMF phase, we have computed $n_F$, the largest value of $n$ for which the groundstate is fully spin polarized, for a variety of ladders, with the results presented in Table \ref{fm}.
An even - odd effect is apparent. For the 3 $\times$ $N$ ladders
$n_F = 0.87$, independent of $N$, while for 4 $\times N$  (as for 2  $\times N$ )
$n_F=0.8$.  For 5- and 6-leg ladders, we were restricted to relatively small $N$, but the trend continues, with $n_F$ slightly greater than 0.8 for the 5-leg and approximately equal to 0.8 for the 6 leg.
Extrapolating either the even or the odd leg ladder results to the 2D limit results in an estimate of $n_F \approx 0.8$ as the ladder width tends to infinity.

We exhibit in Fig. \ref{qt}(b) and \ref{qt}(c), respectively, the ground-state correlations at $n=3/4$ of the longest accessible 4- and 6-leg ladders.  The analogous commensurate checkerboard state we found in the 2-leg ladder is manifest in these correlations, as well.
Indeed, the magnitude of the broken symmetry does not seem to decrease much with increasing ladder width;   the strong (weak) bonds along
legs have typical strengths 0.44 (0.31), $0.38-0.41$ ($0.30-0.31$), and $0.36-0.41$ ($0.31-0.33$) for the 2-, 4-, and 6-leg ladder, respectively.  The corresponding values of the bonds on the strong (weak) rungs of the 4- and 6-leg ladder are 0.38 ($0.32-0.33)$ and $0.37-0.39$ $(0.32-0.33)$, respectively.

For the 4 $\times$ 20 ladder
with $n=3/4$,
$\Delta_c = 0.23 \pm 0.006 t$ and $\Delta_s =0.008 \pm 0.002 t$.
For comparison,  on the 2$\times$20 ladder, the charge gap is of comparable magnitude, $\Delta_c 
=0.286 \pm 0.006t$, but the spin-gap is more than a factor of 20 smaller, $\Delta_s < 0.0003t$. The robustness of the charge gap corroborates the existence of a commensurate insulating checkerboard state.  To the extent that we can think of the low energy spin degrees of freedom as corresponding to a spin 3/2 antiferromagnet defined on the checkerboard lattice, all $(4p+2)$-leg and $4p$-leg ladders correspond, respectively, to $(2p+1)$-leg  and $2p$-leg spin 3/2 Heisenberg ladders, and accordingly 
are expected to be gapless, or to exhibit a spin-gap, albeit one which falls exponentially with increasing $n$ \cite{chakravarty}. Together, these observations strongly support the conclusion that the checkerboard antiferromagnet is the groundstate phase in the 2D limit.

Finally we have found that the groundstates of all the 4-leg ladders we have investigated have total spin $\leq$ 2 for $3/4 > n > 3/5$ and total spin $\leq$ 1 for $3/5 > n$ respectively, corresponding to $M\approx 0$.
We hope to study the apparent Luttinger liquid phase of the 2-leg ladder and its evolution with ladder width toward the
 2D limit using a combination of DMRG and variational treatments, in order to determine whether there is a well defined strong coupling limit\cite{anderson} of the superconducting state that is known to arise at weak coupling \cite{raghu}.

Acknowledgements: We also would like to thank Steve White, Dung-Hai Lee, Fan Yang, George Karakonstantakis, Haijun Zhang, Binghai Yan, and Brian Moritz for helpful discussions. This work was supported, in part, by DOE grant DE-AC02-76SF00515 at Stanford (LL and SAK), DOE grant DE DE-AC02-05CH11231 at Berkeley (HY), NSF grants DMR-0757145 and DMR-0705472 (EB).

\end{document}